# Self-aligned charge read-out for InAs nanowire quantum dots

*Ivan Shorubalko*[*], *Renaud Leturcq, Andreas Pfund, David Tyndall, Roland Krischek, Silke Schön*[†], *and Klaus Ensslin*

Solid State Physics Laboratory, ETH Zurich, 8093 Zurich, Switzerland.

A highly sensitive charge detector is realized for a quantum dot in an InAs nanowire. We have developed a self-aligned etching process to fabricate in a single step a quantum point contact in a two-dimensional electron gas and a quantum dot in an InAs nanowire. The quantum dot is strongly coupled to the underlying point contact which is used as a charge detector. The addition of one electron to the quantum dot leads to a change of the conductance of the charge detector by typically 20%. The charge sensitivity of the detector is used to measure Coulomb diamonds as well as charging events outside the dot. Charge stability diagrams measured by transport through the quantum dot and charge detection merge perfectly.

[*] E-mail: ivan.shorubalko@gmail.com

[†] FIRST lab, ETH Zurich, 8093 Zurich, Switzerland

The control and measurement of individual charges via transport experiments through semiconductor quantum dots (QD) [1] have been complemented by so-called charge detection [2] via a nearby quantum point contact (QPC). These devices are typically realized in two-dimensional electron gases in AlGaAs/GaAs heterostructures. The QPC is placed in close vicinity to the QD giving rise to a strong capacitive coupling between the two systems. The QPC is tuned to a regime where its conductance is a steep function of a gate voltage. This is typically achieved for conductance values below the first plateau [2]. Once an additional electron occupies the QD, the potential in the neighboring QPC is modified by capacitive cross-talk which gives rise to a measurable conductance change. If charge transport is slow enough, which typically happens when the current through the quantum dot is too small to be measured by conventional means, the transport of charges can be monitored in a time-resolved way [3, 4]. The charge sensor allows measurements where conventional transport experiments fail, such as the detection of charge movement between dots [5] or the analysis of shot noise [6] and higher moments of current fluctuations [7].

QDs in InAs nanowires represent another fascinating avenue to realize high quality electronic nanostructures. The bandstructure of InAs gives rise to strong spin-orbit interactions and large confinement energies caused by the small effective mass of the conduction band electrons. This makes QDs in InAs nanowires excellent candidates for spin qubits [8] especially in view of electric field induced spin manipulation [9, 10, 11].

Quantum dots in InAs nanowires have been electrically defined by depositing the nanowires on a predefined gate array [12]. Alternatively top gate fingers can be used to create single and double quantum dots [13]. This way the strength of spin-orbit interactions has been quantified [14, 15] and the suppression of spin relaxation has been detected [16].

By directly growing InP barriers within a InAs nanowire well defined Quantum dots have been realized and investigated in great detail [17]. In a pioneering experiment it has become possible to fabricate a charge read-out [18] on such samples. The nanowires were deposited on a two-dimensional electron gas (2DEG) in an AlGaAs/GaAs heterostructure. Using electron beam lithography a QPC defined by split-gate electrodes was fabricated as close as possible to the quantum dot forming in the nanowire between the InP barriers. This way charging signals of the dot were detected in the transconductance of the QPC. For top-finger-gate defined QDs in a nanowire a neighboring nanowire capacitively connected by a metal strip to the QD has been used as a charge read-out [19].

Here we set out to realize a charge read-out for QDs in InAs nanowires with optimized coupling and sensing ability. An InAs nanowire is deposited on top of a shallow 2DEG. The QD in the InAs nanowire and the QPC in the underlying 2DEG are defined in a single etching step. This way the alignment of the two devices is guaranteed and strong coupling is ensured. We demonstrate a change in the QPC conductance of typically 20% if an additional electron populates the QD.

The sample is an AlGaAs/GaAs heterostructure grown by molecular beam epitaxy and has the following layer structure from the substrate and GaAs buffer: 20 nm $Al_{0.3}Ga_{0.7}As$, 0.6 nm GaAs, Si-delta doping, 1.4 nm GaAs, 2 nm AlAs, 8 nm $Al_{0.3}Ga_{0.7}As$, 5 nm GaAs. This positions the 2DEG interface 37 nm below the sample surface. The 2DEG has a density of $N_s = 4\times10^{11}$ cm$^{-2}$ and mobility $\mu$ = 300'000 cm$^2$/Vs at a temperature of 2 K. The 2DEG is equipped with a mesa structure and ohmic contacts. Next metal-organic vapor phase epitaxy grown InAs nanowires are deposited on the sample surface (similar as in ref. [13]). Using an optical microscope the nanowires with suitable thickness and location with respect to mesa structure and contact pads are identified. As verified by scanning electron microscope (SEM) a typical nanowire diameter of 120 nm is used. Next a 100 nm thick layer of PMMA is exposed by electron beam lithography. The developed areas are etched by freshly mixed $H_2O:H_2SO_4:H_2O_2$ (100:3:1). The etching rate is ~ 1.7 nm/sec, both for the nanowires and for the 2DEG heterostructure. An etching time of 15 seconds has been used for the sample presented here. After removal of the PMMA the sample structure looks as shown in Fig. 1. The etching parameters are carefully optimized such that the constrictions in the nanowire form tunnel barriers suitable for operation of the dot while the etched trenches in the 2DEG underneath are deep enough to act as laterally insulating barriers. Current-voltage characteristics across the 2DEG barriers using e.g. contacts 1 and 2 show that the 2DEG is also pinched off in the trench areas below the quantum wire. The electronic width of the QPC can be adjusted by voltages applied to lateral gates 1 and 3. The QPC itself acts as a gate to tune the electron occupancy in the dot. For the present heterostructure all top gates have a small leakage current. The ohmic contacts A and B to the nanowire are therefore not perfect Schottky contacts with respect to the underlying 2DEG. For the measurements presented in the following we have taken care of this fact by appropriate electronic wiring and biasing, see Fig. 1. Future samples without this leakage problem will allow for even more tunability. All measurements presented here were performed at a temperature of 1.8 K.

Figure 2 (a) shows the current through the dot ($I_{QD}$) for an applied bias of $V_{QD}$ = 1 mV. The potential of the underlying 2DEG containing the QPC ($V_{2DEG}$) serves as one tuning gate. The other gate is the left in-plane gate $V_{LG}$. The ground of the total setup is the left contact of the wire (contact A in Fig. 1) on top of the left lateral gate. Transport through the dot becomes measurable for positive gate voltages where the current signal is larger than our experimental limit of about 10 fA/$\sqrt{Hz}$. Clear Coulomb blockade peaks are observed related to charging of the tunable dot in the nanowire. Parametric rearrangement of charges is also seen. The inset of Fig. 2 (b) shows the current through the QPC ($I_{QPC}$) versus $V_{2DEG}$ for $V_{QPC}$ = 0.2 mV and at $V_{LG}$ = +500 mV. Since the gating of the QPC in this configuration is done by the voltage difference between the nanowire and 2DEG the QPC pinches off for positive $V_{2DEG}$. The transconductance of the QPC $dI_{QPC}/dV_{2DEG}$ is shown in the main panel of Fig. 2 (b). It shows essentially the same features from (a) but is much richer especially in the lower left corner, where the direct dot current is too small to be measured. This general behavior is very similar to what has been observed in GaAs QDs. The bright stripes of strong contrast in Fig. 2 (b) correspond to the steep regions in the QPC characteristics in the inset. An example of this is highlighted by circles. These stripes go from the upper left corner to the lower right corner and originate from the change of the charge on the dot by one electron. The dark lines moving from the lower left corner towards the upper right corner correspond to charging of charge traps in the 2DEG in the vicinity of the QPC. The slope of the QPC conductance is the steepest along the doted line shown in Fig. 2 (b) giving the best charge read-out signal. Thus, further measurements were done tuning both gates ($V_{LG}$ and $V_{2DEG}$) along this line.

Figure 3 shows three panels in which the 2DEG potential is plotted along the horizontal axis. The gate voltage $V_{LG}$ is kept proportional to $V_{2DEG}$ in order to follow the doted line in Fig. 2 (b). Figure 3 (a) shows Coulomb diamonds [1] for the absolute value of the dot current versus bias voltage ($V_{QD}$). The following panel (b) shows the QPC current ($I_{QPC}$) for zero dot bias ($V_{QD} = 0$) and at $V_{QPC} = 0.2$ mV. Pronounced steps of typically 20% of the total QPC current occur each time an additional electron occupies the dot. The total capacitance of the dot is around 16 aF and the leverarm of the QPC serving as a gate is 0.62. For comparison, the typical total capacitance of planar QDs defined in GaAs is around 100 aF and the leverarm of the charge readout is around 0.2. Figure 3 (c) shows the transconductance of the QPC as a function of dot bias. The features with strong contrast having negative transconductance resemble states in the dot with strong coupling. Negative/positive slopes correspond to strong coupling to the source/drain contact. A perfect matching between the QD transport measurements and the detector signal is observed. The detector also displays signals (strong contrast features having positive transconductance) from charge rearrangements not taking place in the dot. These charge rearrangements influence both the dot and the detector signal which could be related to additional traps at the GaAs-InAs interface as suggested in [18]. All of these features have been seen in GaAs QDs [20] but with a generally weaker detector signal. In 2DEGs the QD and QPC detector are separated by a tunnel barrier of typically 100 nm in width. Furthermore the confinement along the growth direction is dominant so that both electronic systems (QD and QPC) have a larger extent of the wave function in the plane of the 2DEG than perpendicular to it. In the present case the QPC detector is 37 nm from the edge of the nanowire. Furthermore the extent of the wave functions of both QPC and QD in the direction perpendicular to the coupling is larger. Therefore it is expected that the electronic coupling between QD and QPC is enhanced compared to the conventional case realized in 2DEGs. This enhanced coupling leads to a larger signal to noise of the QPC detector signal which can be exploited for higher bandwidth charge detection for example for noise measurements and higher order current correlations [7]. Based on this self-aligned etching process one can envision to deposit gate electrodes in the etched grooves to enhance the electrical tunability of both QD and QPC in order to also be able to study several coupled dots. In addition to the previously investigated InAs nanowire quantum dots, which were defined by InP barriers [17] or by gate electrodes [12, 13] our method shows that high-quality quantum dots can also be fabricated by wet etching.

We have presented a self-aligned fabrication process which facilitates a QD in an InAs nanowire being strongly electrostatically coupled to a nearby QPC charge detector in a two-dimensional electron gas. For each additional electron entering the quantum dot the QPC current changes by about 20%. This extraordinary strong signal is useful for time-resolved measurements and for studies of back action between the two quantum devices.

We thank S. Gustavsson for discussions and experimental assistance. Financial support from the Swiss Science Foundation (Schweizerischer Nationalfonds) and ETH Zurich is gratefully acknowledged.

FIGURES:

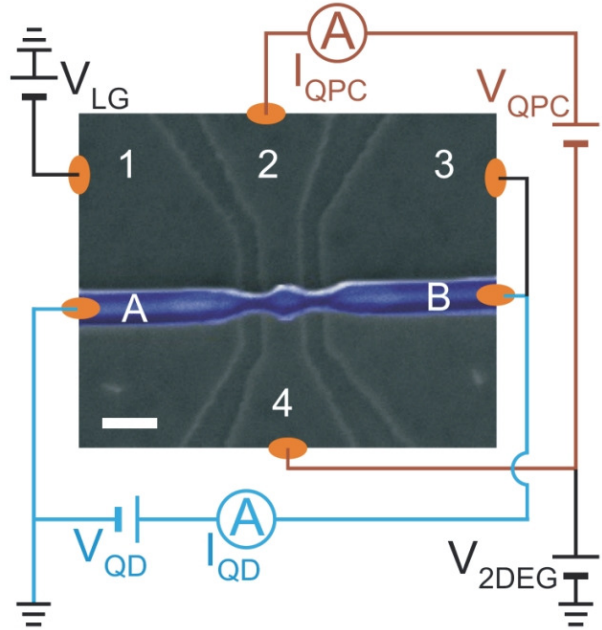

Figure 1.

Scanning electron microscope image of the etched sample structure. The white bar has a length of 200 nm. The top nanowire (colored blue between A and B) is reduced in width in the center where the QD forms between two barriers. The dark lines define insulating barriers in the underlying electron gas. Contacts A and B are source and drain contacts of the nanowire. Contacts 2 and 4 are source drain contacts of the QPC. Contacts 1 and 3 are the lateral gates to tune the electronic width of the QPC. A layout of the measurement circuit with definition of all voltages and currents is sketched around the SEM image.

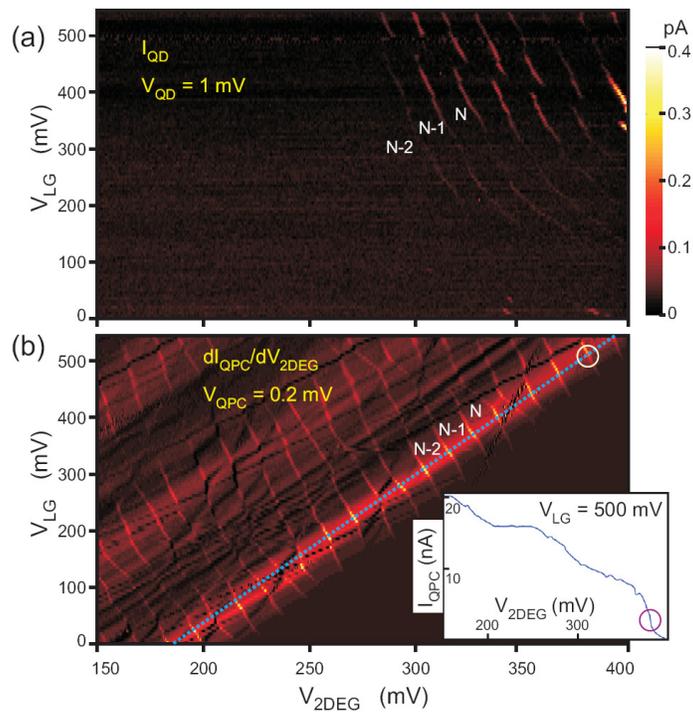

Figure 2.

(a) Current through the quantum dot versus 2DEG potential ($V_{2DEG}$) and left in-plane gate ($V_{LG}$). (b) Transconductance $dI_{QPC}/dV_{2DEG}$ of the QPC for the same parameter range as in (a). Inset of (b): example of current through the QPC as a function of 2DEG potential. Number of electrons on the dot denoted by N, N-1, N-2 relates the features on both graphs. Data in (a) and (b) were recorded simultaneously during a single sweep.

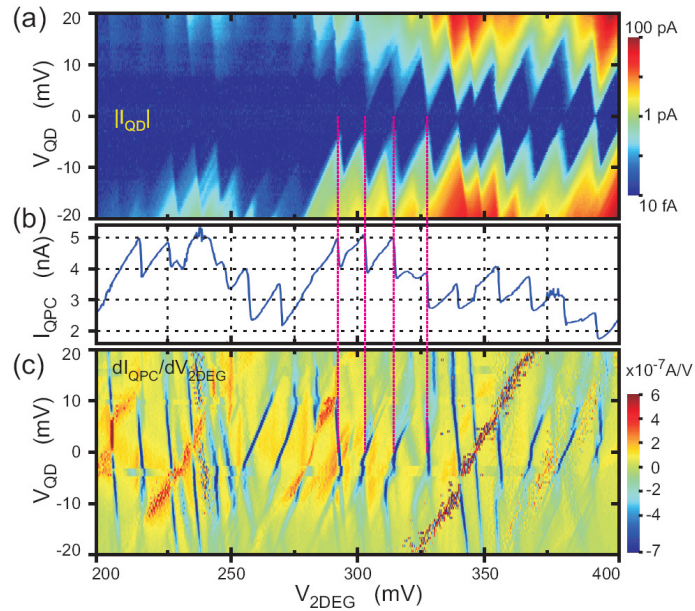

Figure 3.

(a) Color plot of the absolute value of the dot current versus 2DEG potential and bias voltage applied across the dot. From the size of the Coulomb diamonds we estimate a value for the charging energy of typically around 10 meV. (b) QPC current versus 2DEG potential for $V_{QPC} = 0.2$ mV. The steps in current are well aligned with the Coulomb diamonds in (a) and are about 20% of the total current signal. (c) Transconductance of the QPC for the same parameters as in (a). Strong signals with positive/negative slope indicate states which are strongly coupled to source/drain contact. Vertical dashed lines are a guide for the ayes to relate features in the graphs. Data in (a), (b), and (c) were recorded simultaneously during a single sweep.